\documentclass[prd,twocolumn,showpacs,nofootinbib,preprintnumbers]{revtex4}
\usepackage{amsmath}
\usepackage{amsfonts}
\usepackage{graphicx}
\usepackage{hyperref}
\usepackage{mathbbol}
\begin{document}
\preprint{QMUL-PH-06-03}
\title{A Quintessentially Geometric Model}
\date{\today}

\vspace{2cm}

\author{Burin Gumjudpai}
\email{buring@nu.ac.th} \affiliation{Fundamental Physics \&
Cosmology Research Unit, The Tah Poe Academia Institute (TPTP),
Department of Physics, Naresuan University, Phitsanulok 65000,
Thailand}

\author{Tapan Naskar}
\email{tapan@iucaa.ernet.in} \affiliation{IUCAA, Post bag 4,
Ganeshkhind, Pune 411007, India}

\author{John Ward}
\email{j.ward@qmul.ac.uk} \affiliation{Department of Physics,
Queen Mary, University of London, Mile End Road, London, E1 4NS
U.K.}

\begin{abstract}
We consider string inspired cosmology on a solitary $D3$ brane moving in the background of a ring of branes located on a circle of radius $R$.
The motion of the $D3$ brane transverse to the plane of the ring
gives rise to a radion field which can be mapped to a massive non-BPS Born-Infeld type field with a $\cosh$ potential.
For certain bounds of the brane tension we find an inflationary phase is possible, with the string scale relatively
close to the Planck scale. The relevant perturbations and spectral indices are all well within the expected
observational bounds.
The evolution of the universe eventually comes to be dominated by dark energy, which we show is a late
time attractor of the model. However we also find that the equation of state is time dependent, and will
lead to late time Quintessence.
\end{abstract}


\maketitle
\section{Introduction}
It was recently suggested that the rolling open string tachyon, inspired by a class of string theories, can
have important cosmological implications.  The decay of a non-BPS $D3$-brane filling four dimensional
space time leads to a pressureless dust phase which we identify with the closed string vacuum.
The rolling tachyon has an interesting equation of
state whose parameter ranges from $0$ to $-1$. It was therefore thought to be a candidate of inflation
and dark matter, or a model of transient dark energy~\cite{senrev}. However if we rigorously stick to string
theory, the effective tachyon potential contains no free parameter. A viable inflationary scenario
should lead to enough number of $e-$folding, and the correct level of density perturbations. The latter
requires a free parameter in the effective potential which could be tuned to give rise to an
adequate amount of primordial
density perturbations. One also requires an adjustable free parameter in
the effective potential to account for the late time acceleration.

Recently a time dependent configuration in a string theory was investigated and was shown
to have interesting cosmological application \cite{kutasov}. In this scenario a BPS $D3$-brane is placed in the
background of several coincident, static $NS5$-branes which are extremely heavy compared to the $D3$-brane and form
an infinite throat in the space time. This system is inherently non-supersymmetric because the two different kinds of
branes preserve different halves of the bulk supersymmetries.
As a result the $D3$ brane can be regarded as a probe of the warped background and
is gravitationally attracted toward the $NS5$-branes. Furthermore there exists an exact conformal field theory
description of this background where the number of five-branes determines the level of the WZW current algebra
\cite{saha}, which allows for exact string based calculations.
Despite the fact that the string coupling diverges as we approach the fivebranes, it was shown that we
can trust our effective Dirac-Born-Infeld (DBI) action to late times in the evolution provided that the energy
of the probe brane is sufficiently high. In any event, as the probe $D3$-brane approaches the background branes
the spatial components of the energy-momentum tensor tend to zero in exactly the same way as in
the effective action description of the open string tachyon. Thus it was anticipated that the dynamics of branes
in these backgrounds had remarkably similar properties to rolling tachyon solutions.
This relationship was further developed by Kutasov who showed that it was possible to mimic the open string
tachyon potential by considering brane motion in a specific kind of 10D geometry. In order to do this one
must take the action of the BPS probe brane in the gravitational background and map it to a non-trivial scalar
field solution described by the non-BPS action \cite{DBI}. The new field is essentially a holographic field
living on the world-volume of the brane, but encodes all the physics of the bulk background.
This is known as the geometrical tachyon construction.
Another particularly interesting solution considered the background branes distributed around a ring of radius $R$,
which was analysed in ref \cite{TW1, TW2}, and whose geometry is described by a coset model \cite{israel},
again potentially opening the way for an exact string calculation.

It seems natural to enquire as to whether these geometrical tachyon solutions have any relevance for cosmology,
since they neatly avoid the problems associated with open string tachyon inflation \cite{KL} by having
a significantly different mass scale. This change in scale is due to the motion of the probe brane in a gravitationally
warped background, provided by the branes in the bulk geometry. In essence, this is an alternative formulation of the
simple Randall-Sundrum model \cite{randall}. More recently, flux compactification has opened up the possibility of realising these models
in a purely four-dimensional string theory context \cite{warpcosm}. The fluxes form a throat which is glued onto a compact manifold
in the UV end of the geometry. The warp factor in the metric has explicit dependence on the fluxes, and so provides
us with a varying energy scale. The recent approaches to brane cosmology \cite{BA} are based on the motion of $D3$-branes in
these compactifications. Typically we find $\bar{D}3$-branes located at some point in the IR end of the throat, which
provide a potential for a solitary probe brane, with the inflaton being the inter-brane distance. In this context
we can obtain slow roll inflation, and also the so-called DBI inflation \cite{dbiinflation}, which relies heavily on the red-shifting of
energy scales. However flux compactification models have an unacceptably large number of vacua, characterised by the
string landscape. They are also low energy models, where the string scale is significantly lower than the Planck scale and
so there is no attempt to deal with the initial singularity. In addition, we require multiple throats attached to the
compact manifold where the standard model is supposed to live, however there is no explanation for the
decoupling of the inflaton sector. These problems need to be addressed if we are to
fully understand early universe cosmology in a string theory context.
The alternative approach is to consider cosmology in the full ten dimensional string theory. Although these models
are plagued by their own problems there is a definite sense of where the standard model is assumed to live, and a natural realisation
of inflation. Furthermore we can invoke a Brandenberger-Vafa type mechanism to explain the origin of our $D3$-brane,
arising from the mutual cascade annihilation of a gas of $D9$-$\bar{D}9$-branes \cite{BVmechanism}.

An alternative approach is compactify our theory on a compact manifold, where some mechanism is employed to
stabilise the various moduli fields. This will naturally induce an Einstein-Hilbert term into the four dimensional action
\cite{verlinde}. However this is a highly non-trivial problem whose precise details remain unknown. Despite being unable
to embed this into String Theory, we can still learn a great deal about the physics of the model - as emphasised by recent works \cite{NS5cos}.

A specific case of interest has been to study inflation in the ring solution ref \cite{TWinf}. Due to the unusual nature
of the harmonic function we find decoupled scalar modes, one transverse to the ring plane and the other
inside the ring. The cosmology of modes inside the ring have been studied in ref \cite{ringinf}.
In this note we will consider the situation in which the $D3$-brane moves in the transverse direction to the ring.
Performing the tachyon map in this instance yields a $\cosh$ type potential implying that the resulting scalar
field in the dual picture is massive.
It is interesting that in this setting we do not have to worry about
the continuity condition around the ring. And unlike the longitudinal motion, we have an
analytic expression for the effective potential every where in the transverse directions.
We study the cosmological application of the resulting scenario and show that the model leads
to an ever accelerating universe. We study the autonomous form of field evolution equation in the
presence of matter and radiation and show that the de-Sitter solution is a late time attractor of the
model. We also demonstrate the viability of the geometrical tachyon for dark energy in the setting under consideration,
arising in a natural way due to the non-linearity of the DBI action.
In the next section we will introduce the string theory inspired model, and discuss how we an relate it to
four dimensional cosmology. In section III we will consider the more phenomenological aspects of our model by comparing
our results with experimental observation. Section IV shows how we have a natural realisation of reheating in our model,
whilst section V discusses the final stage of dark energy domination. Our model predicts that the equation of state parameter will
tend to $\omega \sim -1$, but on even larger timescales we expect it to increase toward zero as in models of quintessence \cite{quintessence}.
We will conclude with some remarks and a discussion of possible future directions.
%%%%%%%%%%%%%%%%%%%%%%%%%%%%%%%%%%%%%%%%%%%%%%%%%%%%%%%%%%%%%%%%%%%%%

\section{Geometrical Scalar Field and Coupling to Gravity.}
We begin with the string frame CHS solution for $k$ parallel, static $NS5$ branes in type IIB String Theory \cite{Callan, sfet}.
The metric is given by:
\begin{eqnarray}
ds^2 &=& \eta_{\mu \nu}dx^{\mu}dx^{\nu}+F(x^n)dx^mdx^m,
\end{eqnarray}
where $\chi$ is the dilaton field define as $e^{2(\chi-\chi_0)} = F(x^n)$, and there exists the three
form field strength of the NS B-field $H_{mnp}=-\varepsilon^q_{mnp}\partial_q \phi $. Here $F(x^n)$ is the
harmonic function describing the position of branes. For a large
number of branes we can consider the throat approximation, which amounts to dropping the factor
of unity in the function. Inherently we are decoupling Minkowski space time from the theory, and therefore
only interested in the region around the $NS5$-branes. The harmonic function is given by:
\begin{eqnarray}
F &=& 1+\frac{kl_s^2~\sinh(ky)}{2R\rho ~ \sinh(ky)~(\cosh(ky)-\cos(k\theta))} \nonumber \\
  &\approx &\frac{kl_s^2~\sinh(ky)}{2R\rho ~ \sinh(ky)~(\cosh(ky)-\cos(k\theta))},
\end{eqnarray}
where $\rho$, $\theta$ parameterise polar coordinates in the ring plane, and the factor $y$ is given by:
\begin{eqnarray}
{\rm cosh}(y) &=& \frac{R^2+\rho^2}{2 R \rho}.
\end{eqnarray}
We put a probe $D3$ brane at the centre of $NS5$ branes, as mention in the introduction this brane will
move toward the circumference due to gravitational interaction if it shifted a little
from the centre keeping the brane in the plane of the ring; the cosmology in this case
is described elsewhere. We consider the case where the probe brane lies in the centre
of the ring but shifted a little from the plane. In this case the probe brane shows
transverse motion. Note that because of the form of the DBI action, the configuration here is actually $S$-dual
to the $D5$-brane ring solution. The only difference is the shift of $k \to 2g_sk$ in the harmonic function.
The physics however are very different as we know that $F$-strings cannot end on the $NS5$-branes, but can
end on the $D5$-branes. This implies that in the case of the $D5$-brane ring we can have additional open
string tachyonic modes once the probe brane starts to resolve distances of order of the string scale.
The cosmological implications for this extra field were discussed in \cite{NS5cos}.

For the brane at the center ($\rho=0$) moving transverse to the ring ($\dot{\rho}=0$), the harmonic
function is given by:
\begin{eqnarray}
F(\sigma) &=& \frac{kl_s^2}{R^2+\sigma^2},
\end{eqnarray}
and the DBI action for the probe brane can be written in the following form, in static gauge
\begin{equation}
S = -\tau_3 \int d^4 \xi \sqrt{F^{-1}-\dot{\sigma}^2}.
\end{equation}
The tachyon map in this instance arises via field redefinition. We define the following
scalar field, which has dimensions of length
\begin{equation}
\phi(\sigma) = \int \sqrt{F}d\sigma,
\end{equation}
which maps the BPS action to a form commonly used in the non-BPS case \cite{DBI}
\begin{equation}
S = -\int d^4 \xi V(\phi) \sqrt{1-\dot{\phi}^2},
\end{equation}
where $V(\phi)$ is the potential for the scalar field which describes the changing tension of the
$D$-brane.

From the above mapping we get the solution of field as:
\begin{eqnarray}
\phi(\sigma) &=& \int^{\sigma}_{0}\sqrt{F(\sigma')}d\sigma'\nonumber \\
           &=& \sqrt{kl_s^2}\ln\left(\frac{\sigma}{R}+\sqrt{1+\frac{\sigma^2}{R^2}}\right) \nonumber \\
           &=& \sqrt{kl_s^2} {\rm arcsinh} \left(\frac{\sigma}{R}\right) \\
V(\phi)    &=& \frac{\tau_3}{\sqrt{F}} \nonumber \\
           &=& \frac{\tau_3 R}{\sqrt{kl_s^2}}\cosh\left(\frac{\phi}{\sqrt{kl_s^2}}\right) \label{poten}.
\end{eqnarray}
Clearly we see that $\phi \to \pm \infty$ as $\sigma \to \pm \infty$, and that at the minimum of the potential
we have $\phi=0$ \footnote{We must bear in mind that our approximation of the harmonic function prevents us from
taking the $\sigma \to \infty$ limit.}.
The potential of the field suggests that the mass is given by $1/kl_s^2$, corresponding to a massive scalar
fluctuation.
One may ask if there is a known string mode exhibiting this profile. In fact the fluctuations of a massive scalar
were computed in \cite{garousi} using a similar approach to the construction of the open string tachyon mode
in boundary conformal field theory \cite{senrev}.
This field was then used in ref. \cite{GST, scalarfield} as a candidate for the inflaton living on a $\bar{D}3$-brane
in the KKLT scenario ref. \cite{KKLT}. The potential for the scalar is
known to fourth order and was been assumed to be exponential in profile, although globally it may be hyperbolic.

In order to discuss the cosmological evolution of our scalar field
we need to couple our effective action to four dimensional
Einstein gravity. There are several ways we can accomplish this.
Firstly we can consider the Mirage Cosmology scenario
\cite{mirage}. This requires us to re-write the induced metric on
the $D3$-brane world-volume in a Friedmann-Robertson-Walker (FRW)
form. The universe will automatically be flat, or closed if we
imagine the $D$-brane to be spherical. The problem here is that
there is no natural way to couple gravity to the brane action and
therefore we must insert it by hand, however the cosmological
dynamics are expected to be reliable virtually all the way to the
string scale. The second option is a slight modification of the
first. We imagine that the bulk is infinite in extent, and that
the $D3$-brane is again coupled to gravity through some unknown
mechanism. However rather than writing the induced metric in FRW
form, we switch to the holographic theory. Now, the tachyon
mapping in this case is only concerned with time-dependent
quantities, and in particular only with the temporal component of
the Minkowski metric. Therefore we choose to include a scale
factor component in the spatial directions. This means that we
have a cosmological coupling for the holographic scalar field, and
the universe lives on the $D3$-brane world-volume. The final
approach would be to compactify the theory down to four
dimensions. In order to do this we need to truncate the background
to ensure the space is compact \cite{warpcosm}. In our case the
ring can naturally impose a cut-off in the planar direction,
however we must still impose some constraint in the transverse
direction to the ring plane. Our solution simplifies somewhat if
we can consider the $R \to 0$ limit, or equivalently the $\sigma
\gg 1$ limit, as the background will appear point like. Smoothly
gluing the truncated space to a proper compact manifold will now
automatically include an Einstein-Hilbert term in the effective
action \cite{verlinde}. However, although we now have a natural
coupling to gravity, the compactification itself is far from
trivial as we also need to wrap two of the world-volume directions
of the $NS$5-branes on a compact cycle. In order to proceed we
must first uplift the full solution to M-theory \footnote{This was
discussed by Ghodsi et al in \cite{NS5cos}. We refer the
interested reader there for more details.}, where we now have a
ring of $M5$-branes magnetically charged under the three-form
$C_{(3)}$. Compactification demands that the magnetic directions
of the three-form are wrapped on toroidal cycles, which is further
complicated by the ring geometry and will generally result in
large corrections to the potential once reduced down to
four-dimensions. So, although we have a natural gravitational
coupling we may have large corrections to the theory. The complete
description of this compactification is interesting, but well
beyond the scope of this note and should be tackled as a future
problem. However we could also assume a large volume toroidal
compactification, where again all the relevant moduli have been
stabilised. Provided we introduce some 'sink' for the five-brane
charge, located at the some distant point in the compact space,
and also only concentrate on the region close to the branes so
that the harmonic function remains valid and we will have an induced
gravitational coupling in the low energy theory. The corrections
to the scalar potential in this region of moduli space may well be
sub-leading with respect to the scalar field dynamics and thus we
can treat our model as the leading order behaviour.
\\
Recent work in this direction has been concerned with the compactification approach \cite{NS5cos, TWinf}, where it was assumed
all the relevant moduli are fixed along the lines of the KKLT model \cite{KKLT} and that all corrections to the
potential are sub dominant. We will tentatively assume that this will also hold in our toy model.
\\
We can now analyse our four dimensional minimally coupled action, where we find the following solutions to the
Einstein equations
\begin{eqnarray}
H^2 &=& \frac{V(\phi)}{3M_p^2\sqrt{1-\dot{\phi}^2}} \\
\frac{\ddot{a}}{a}&=&\frac{V(\phi)}{3M_p^2\sqrt{1-\dot{\phi}^2}}\left(1-\frac{3\dot{\phi}^2}{2} \right).
\end{eqnarray}
These expressions are different to those associated with a traditional canonical scalar field. In particular we see that
inflation will automatically end once $\dot{\phi}^2 \sim 2/3$ as in the tachyon cosmology models \cite{tachinfl, tachyonpapers, sen}.
For completeness we write the
equation of motion for the inflaton derived from the non-BPS action as follows
\begin{equation}\label{eq:eom}
\frac{V(\phi) \ddot{\phi}}{1-\dot{\phi}^2} + 3HV(\phi)\dot\phi + V'(\phi)=0,
\end{equation}
where dots are derivatives with respect to time and primes are derivatives with respect to the field. Note that we are
suppressing all delta functions in the expressions.
We can now proceed with the analysis of our theory in the usual manner. It must be noted that this
model corresponds to large field inflation, where the initial value of the scalar field must satisfy
the following condition
\begin{equation}
\phi_0 \ll \sqrt{kl_s^2} {\rm arccosh}
\left(\frac{\sqrt{kl_s^2}}{R} \right),
\end{equation}
according to our truncation of the harmonic function.

Note that in what follows we will frequently switch between the field theory and the bulk geometry.
The latter is more geometrical and so provides us with extra intuition about the physics of the solution, however both are
equivalent - at least in this simplified model.

Using the slow-roll approximation, $H^2 \simeq V(\phi)/3M_p^2$ and $3H \dot{\phi} \simeq -V_{\phi}/V$,
the e-folding %N=\ln a$ is:
\begin{eqnarray}
N &=&\int _t ^{t_f} H dt \nonumber \\
   &=& \frac{\tau_3R\sqrt{kl_s^2}}{M_p^2}\int_{x(\phi_f)}^{x(\phi)}\frac{\cosh^2x}{{\rm sinh}\, x}dx \nonumber \\
  &=& s\left[-\cosh(x_f)+\cosh(x)-\ln\left(\frac{{\tanh(x_f/2)}}{{\tanh(x/2)}}\right)\right]. \nonumber \\
\end{eqnarray}
Where we have introduced the dimensionless quantities $x = \phi/{\sqrt{kl_s^2}}$ and $s = {\tau_3R\sqrt{kl_s^2}}/{M_p^2}$.\\
Further defining the new quantity: $ y \equiv \cosh x $ we can write the number of e-folds as follows:
\begin{eqnarray}
N &=& s\left[-y_f+y-\frac{1}{2}\ln\left(\frac{(y_f-1)(y+1)}{(y_f+1)(y-1)}\right)\right]
\label{N}
\end{eqnarray}
Now, the relevant slow-roll parameter is defined as $\epsilon \equiv -\dot{H}/H $
which in our solution reduces to
\begin{eqnarray}
\epsilon &=& \frac{y^2-1}{2sy^3}.
\end{eqnarray}
Note that our model is explicitly non-supersymmetric, and therefore we don't need to calculate the second slow roll parameter $\eta$ since we anticipate that this will be trivially satisfied if $\epsilon$ is.
At the end of inflation $\epsilon =1$, then $y_f \equiv f(s) $ is given by the root of above
equation, setting $\epsilon=1$
\begin{eqnarray}
f(s) &=& \frac{1}{6s}\left[g(s)+\frac{1}{g(s)}+1\right]
\end{eqnarray}
where $g(s) = \left(-54s^2+1+6s\sqrt{3(27s^2-1)}\right)^{1/3}$
From eqn(\ref{N}) the equation for $y$ is:
\begin{eqnarray}
\ln\left(\frac{y+1}{y-1}\right)-2y &=& -\frac{2N}{s}-2f(s)-\ln\left(\frac{f(s)-1}{f(s)+1}\right)\nonumber \\
\label{transc}
\end{eqnarray}
For $s>1$ and as $y_{\rm min}=1$, $\epsilon$ always remains less
than one leading to an ever accelerating universe. Thus, in this
case the geometrical scalar field in the present setting is not
suitable to describe inflation but can become a possible candidate
of dark energy. However if $\tau_3$ is small enough so that $s<1$,
then we will find that inflation is possible as the slow roll
parameter will naturally tend toward unity.
There is a critical bound $s \leq 1/(3\sqrt{3})$, which must be satisfied if we are to consider inflation in this context.\\

%%%%%%%%
\section{Inflationary Constraints.}
To know the observational constraint on $s$ we have to calculate the density perturbations. In the slow-roll
approximation, the power spectrum of curvature perturbation is given by \cite{MFB, HN, SV}:
\begin{eqnarray}
P_S &=& \frac{1}{12 \pi^2 M_p^6}\left(\frac{V^2}{V_{\phi}}\right)^2 \nonumber \\
    &=& \frac{\tau_3^2 R^2}{12 \pi^2 M_p^6}\left(\frac{\cosh^2(\phi/\sqrt{kl_s^2})}{\sinh(\phi/\sqrt{kl_s^2})}\right)^2
\label{pert}
\end{eqnarray}
The COBE normalisation corresponds to $P_S \simeq 2 \times 10^{-9}$ for modes which crossed $N=60$ before
the end of inflation \cite{constraints} which gives the following constraint:
\begin{eqnarray}
k(l_sM_p)^2 &\simeq & \frac{10^9}{12\pi^2}\frac{s^2\cosh^4(\phi/\sqrt{kl_s^2})}{\cosh^2(\phi/\sqrt{kl_s^2})-1}
\label{cons}
\end{eqnarray}
From the numerics using eqn(\ref{transc}) and eqn(\ref{pert}), we find that
\begin{eqnarray}
k(l_sM_p)^2 \geq 3 \times 10^{10}
\end{eqnarray}
which corresponds to $s \sim 10^{-3}$ when we impose the constraints $\tau_3 = 10^{-10}M_p^4$ and $R=10^2/M_p$ which
we regard as being typical values. The constraint on the tension in fact implies the following relationship
\begin{equation}
\frac{M_p}{M_s} \sim \frac{10^2}{g_s^{1/4}},
\end{equation}
which we need to be consistently satisfied.
However, note that because of our basic assumptions about the theory we will generally obtain the bound
\begin{equation}
\frac{\tau_3 R}{M_p^3} \le \frac{1}{9 \times 10^5}.
\end{equation}
If we write the tension of the brane in terms of fundamental parameters we can estimate the relationship between the
String and Planck scales using the fact that we require $R > M_s^{-1}$ for the action to be valid
\begin{equation}
\frac{M_p}{M_s} \ge \frac{15}{g_s^{1/3}},
\end{equation}
where $g_s$ is the string coupling constant. Note that this potentially constrains the String scale to be close to the
Planck scale, as even if we demand weak coupling with $g_s = 0.001$ this gives us $M_p \ge 10^2 M_s$. Of course this
is only a bound, and in our model we are treating this as a free parameter. In any event our typical values
are consistent and thus we feel free to proceed. We should note that from a string theoretic point of view we should not take $s$ as being a variable
in this model. However our earlier analysis has shown that if we wish to consider non-eternal inflation, there exists a maximum bound on this parameter
which is quite small. Thus we can make the assumption that $s$ will always be small, with appropriate tuning of the ratio of the string and Planck scales..
In the following analysis we will always be assuming that this is satisfied so as to avoid en eternal inflation scenario. Of course, in the string theory picture
we have a probe brane moving in a non-trivial background geometry, and we would expect that the $RR$ charge on the brane will be radiated away in the form of
closed string modes. This effectively means that there is an additional decay constant in the definition of the field $\phi$, which we have neglected in this note. Thus
what we have here is a first-order approximation to the behaviour of the solution. It remains an open question as to whether we can define a tachyon map in this
instance - and how this changes the inflationary scenario described here.

At leading order in our solutions, where $s$ is assumed to be small and making sure our effective action remains valid, we obtain
\begin{eqnarray}
k(l_sM_p^2)^2 &\simeq& \frac{10^9}{48\pi^2}\left(2N+1\right)^2
\label{l_bound_kl}
\end{eqnarray}
which corresponds to $s \sim 10^{-5}(2N+1)$ and $y\sim \frac{(2N+1)}{2s}$,
when $\tau_3 = 10^{-10}M_p^4$ and $R=10^2/M_p$.
Again, more generally we would find the following upper limit on the solution
\begin{equation}
s \le 10^{-3} (2N+1),
\end{equation}
which is easily satisfied by our typical values. In fact our results remain robust when compared to the WMAPII and SDSS results combined \cite{wmap2}.
The new data constrains $n_s = 0.98 \pm 0.02$ at the $68$ confidence level, and $r < 0.24$ at the $95$ confidence level.

The spectral index of scalar perturbations is defined as \cite{MFB, HN, SV}:
\begin{eqnarray}
n_S-1 &\equiv& -4\frac{M_p^2V^2_{\phi}}{V^3}+2\frac{M_p^2V_{\phi \phi}}{V^2} \nonumber\\
     &=&\frac{2}{s}\left(\frac{2-y^2}{y^3}\right)
\label{n_s}
\end{eqnarray}
The spectral index of tensor perturbations is defined as:
\begin{eqnarray}
n_T &=& -\frac{M_p^2V_{\phi}}{V^3} \nonumber \\
    &=& -\frac{1}{s}\left(\frac{y^2-1}{y^3}\right)
\label{n_T}
\end{eqnarray}
The tensor-to-scalar ratio is:
\begin{eqnarray}
r &\equiv& 8\frac{M_p^2V^2_{\phi}}{V^3} \nonumber \\
  &=&\frac{8}{s}\left(\frac{y^2-1}{y^3}\right)
\label{r}
\end{eqnarray}
With the limit $s\to 0$ we get
\begin{equation}
\begin{array}{lll}
n_S = 1-\frac{4}{(2N+1)}, & n_T = -\frac{2}{(2N+1)}, & r=\frac{16}{(2N+1)}
\end{array}
\end{equation}
For $N=60$, we get $n_S=0.96694$ and $r= 0.13223$; for $N=50$, we get $n_S=0.96040$ and
$r=0.15842$. We know from observations that the constraint on the tensor-to-scalar
ratio is $r<0.36$ \cite{obcon, obcon2}, and so our model appears to be well within this bound.
%%%%%%%%%%%%%%%%%%%%%%%%%%%%%%%%%%%%%%%%%%%%%%%%%%%%%%%%%%%%%%%%%%%%%%%%%%%%%%%
\section{Reheating}
We see that the potential is a symmetric potential with a
minima. In terms of the bulk field $\sigma$ it can be written as:
\begin{eqnarray}
V(\sigma) &=&\frac{\tau _3 R}{2\sqrt{kl_s^2}}\left[\left(\frac{\sigma}{R}+\sqrt{1+\frac{\sigma ^2}{R^2}}\right)\right.\nonumber \\
           && \left.+\left(\frac{\sigma}{R}+\sqrt{1+\frac{\sigma ^2}{R^2}}\right)^{-1}\right]
\end{eqnarray}
Now the question is would the brane oscillate back and forth through the ring, and if so what are the necessary
conditions for oscillation? In the bulk picture we would naturally anticipate oscillation with a decaying amplitude
due to $RR$-emission. Moreover the minimum of the potential in this case is actually metastable. 
However this has not been verified as we need to calculate the energy emission in the coset model
description \cite{israel}, which we leave as future work. This will alter the dynamics of the inflaton field as discussed in the previous section.
\\
In any event we may also expect similar behaviour once our field
is coupled to gravity, with the damping being provided by the
Hubble term. This is particularly important because we may find
inflation occurring in the phase space region beyond $s \ge s_{\rm
crit}$, once enough damping has occurred. The relevant dynamical
equations are the inflaton field equation (\ref{eq:eom}) and the
Friedmann equation. We repeat them below for convenience.
\begin{eqnarray}
\ddot{\phi}+3H\dot{\phi}(1-\dot{\phi}^2) +
\frac{V_{\phi}}{V}(1-\dot{\phi}^2) &=&0 \label{dyn_eqn}
\end{eqnarray}
\begin{eqnarray}
H^2 &=& \frac{1}{3M_p^2}\left(\frac{V}{\sqrt{1-\dot{\phi}^2}}+\rho _B\right)
\end{eqnarray}
where the terms inside the curly brackets cause damping. For an
easy treatment let us first consider the slow-roll approximation,
then in the damping equation only the $\dot{\phi}$ term remains
and all other powers of $\dot{\phi}$ can be ignored. That is to
say we are considering the case near the stable point. Then $H^2
\sim \frac{\tau _3R}{3M_p^2\sqrt{kl_s^2}}+\frac{\rho_B}{3M_p^3}$
is constant and $\frac{V_{\phi}}{V} \sim \frac{2}{kl_s^2}\phi$.
The equation of motion is then:
\begin{eqnarray}
\ddot{\phi} + 3\dot{\phi}
\sqrt{\left(\frac{\tau_3R}{3M_p^2\sqrt{kl_s^2}}+\frac{\rho_B}{3M_p^2}
\right)} + \frac{2}{kl_s^2} \phi &=& 0 \label{phi_ddot}
\end{eqnarray}
for critically damped motion we need:
\begin{eqnarray}
\left(\frac{\tau_3
R}{3M_p^2\sqrt{kl_s^2}}+\frac{\rho_B}{3M_p^2}\right) &=&
\frac{8}{9kl_s^2} \label{damping_condition}
\end{eqnarray}
If the RHS of (\ref{damping_condition}) is greater than the LHS we will find oscillations but it is reduced by damping which depends on the size of the damping factor ($=\frac32\sqrt{\left(\frac{\tau _3R}{3M_p^2\sqrt{kl_s^2}}+\frac{\rho_B}{3M_p^2}\right)}$), compared to the
oscillation frequency ($=\sqrt{\frac{8}{kl_s^2}-9\left(\frac{\tau _3R}{3M_p^2\sqrt{kl_s^2}}+\frac{\rho_B}{3M_p^2}\right) }$).\\
From the definition of $\Omega_B$ setting it to $0.3$, we get $\rho _B = \frac{3\tau _3 R}{7\sqrt{kl_s^2}}$, then from
eqn(\ref{damping_condition}) we obtain
\begin{eqnarray}
%\frac{10\tau _3R}{21M_p^2}\sqrt{kl_s^2} &>& \frac89 ~~\textrm{Over damp} \nonumber \\
%                                        &=& \frac89 ~~\textrm{Critically damp} \nonumber \\
%                                        &<& \frac89 ~~\textrm{Oscillatory with decaying amplitude.} \nonumber \\
s &>& \frac{168}{90} ~~\textrm{Over damped} \nonumber \\
&=& \frac{168}{90} ~~\textrm{Critically damped} \nonumber \\
&<& \frac{168}{90} ~~\textrm{Oscillatory with a decaying amplitude.} \nonumber \\
\end{eqnarray}
Recall from the previous section that for us to have non-eternal inflation there is a maximal bound for $s$, and so only the
last solution can be considered physical.
From the constraint we get $\sqrt{kl_s^2} \sim 10^5M_p^{-1}$,
$\tau _3 \sim 10^{-10}M_p^4$ and $R \sim 10^2M_p^{-1}$. Hence it
is oscillatory near the critical point. The energy of the decaying
scalar field is used in expansion and particle production. If the
rate of expansion of universe is much less than the decaying rate
of the amplitude of the field then most of the energy released by
the scalar field goes to reheating. The explicit solution of eqn
(\ref{phi_ddot}) is:
\begin{eqnarray}
\phi(t) & &
=\phi_0e^{\left[-\frac32t\sqrt{\frac{10\tau_3R}{21M_p^2\sqrt{kl_s^2}}}\right]}
e^{\left[\pm \mathbb{i}
t\sqrt{\frac{2}{kl_s^2}-\frac{15\tau_3R}{14M_p^2\sqrt{kl_s^2}}}\right]}
\end{eqnarray}
The ratio of rate of field decay to the rate of expansion of universe is is defined to be:
\begin{eqnarray}
\Theta &\equiv |\frac{\dot{\phi}}{H\phi}|
\end{eqnarray}
For this case we find:
\begin{eqnarray}
\Theta &=& \sqrt{\frac{21M_p^2}{5\tau _3 R\sqrt{kl_s^2}}}
\end{eqnarray}
The above quantity can be made to be less than one by adjusting the various parameters.\\
Using eqn(\ref{l_bound_kl}) we obtain:
\begin{eqnarray}
%\Theta &=& \frac{537.66}{\sqrt{2N+1}}
\Theta &\sim& \sqrt{\frac{21 \times 10^5}{5(2N+1)}}
\end{eqnarray}
which allows us to write the parameter as a function of the number of e-foldings, provided we can trust our small $s$ expansion.
We know that reheating ends when $\Theta =1$, thus the minimal number of e-foldings we require for this to be satisfied is
\begin{eqnarray}
N_{\rm end} &\sim& 10^{5}.
\end{eqnarray}
Clearly this is a large number of e-foldings, and this should motivate us to do a more thorough analysis. For now it would
appear that unless there is a large amount of fine tuning, reheating would not end in this scenario.
The difficulty is that we cannot use the WKB approximation in this case due to rapid fluctuations
in the variation of the potential. Moreover, the analysis will be incomplete without specifying
the exact form the gravitational coupling - as there will be corrections to the effective action
arising from any compactification. For these reasons we will postpone the analysis and return
to it in a later publication.

%%%%%%%%%%%%%%%%%%%%%%%%%%%%%%%%%%%%%%%%%%%%%%%%%%%%%%%%%%%%%%%%%%%%%%%%%%%%%%%
\section{Dark energy}
What are the implications of our model for dark energy\footnote{See \cite{darkrev} for an excellent review.}? It is well known that the non-linear form of the DBI action
admits an unusual equation of state, which is of the form
\begin{eqnarray}
\omega &=& \frac{P}{\rho} \\
&=& \dot{\phi}^2-1 \nonumber
\end{eqnarray}
where $P$ and $\rho$ are the pressure and energy densities respectively.
In tachyon models the field is moving relativistically near the vacuum and the equation of state will
tend to $\omega \sim 0$, which is problematic for reheating.
However our model
has significantly different late time behaviour because our scalar field will oscillate about the minimum of its
potential, eventually coming to a halt at the minimum.
Therefore we expect the equation of state to become $\omega \sim -1$, corresponding to the vacuum energy of the universe.
This motivates us to analyse our system as a potential candidate for dark matter.
One problem, however, is that the reheating phase doesn't seem to have a natural termination point. Rather, reheating of the universe
continues whilst the brane oscillates around the minimum of the potential, and then terminates in what appears to be a dark energy dominated phase.
From the perspective of model building this is obviously a difficult problem. For now let us assume that there is some ad hoc mechanism which
ends inflation, and look at the evolution of the system in this dark matter dominated phase.
The corresponding evolution equations of interest are:
\begin{eqnarray}
\frac{\ddot{\phi}}{1-\dot{\phi^2}}+3H\dot{\phi} + \frac{V_{\phi}}{V} &=& 0 \label{phidot}\\
\dot{H}+\frac{V(\phi)\dot{\phi^2}}{2M_p^2\sqrt{1-\dot{\phi^2}}}+\frac{\gamma \rho_B}{2M_p^2} &=& 0
\end{eqnarray}
where we have included contribution from a barotropic fluid in the second equation.
Defining the following dimensionless quantities:
\begin{eqnarray}
Y_1 &=& \frac{\phi}{\sqrt{kl_s^2}} \nonumber \\
Y_2 &=& \dot{\phi},
\label{dimless}
\end{eqnarray}
and using eqn(\ref{phidot}) and eqn(\ref{dimless}) we get the autonomous equations:
\begin{eqnarray}
Y'_1 &=& \frac{1}{\sqrt{kl_s^2}H}Y_2 \label{y1}\\
Y'_2 &=& -\left(1-Y_2^2\right)\left(3Y_2+\frac{1}{H}\frac{dY_3}{dY_1}\right) \label{y2}
\end{eqnarray}
Where we have switched to using the number of e-folds as the time parameter, and now primes denote derivatives with respect to $N$. The final expressions we require can be read off as
\begin{eqnarray}
Y_3 &=& \ln \left(\frac{V(\phi)}{3M_p^2}\right) \nonumber
\end{eqnarray}

\vspace{0.1in}
%%%%%%%%%%%%%%%%%%%%%%%%%%%%%%%%%%%%%%%%%%%%%%%%%%%%%%%%%%%%%%%%%%%%%%%%%%%%%%%%%%%%%%%%%%%%%%%%%%
%%% Phase Potrait                                                                                 %%
%%%%%%%%%%%%%%%%%%%%%%%%%%%%%%%%%%%%%%%%%%%%%%%%%%%%%%%%%%%%%%%%%%%%%%%%%%%%%%%%%%%%%%%%%%%%%%%%%%
\begin {figure}[h]
\begin{center}
\includegraphics[width=2.5in, angle=-90]{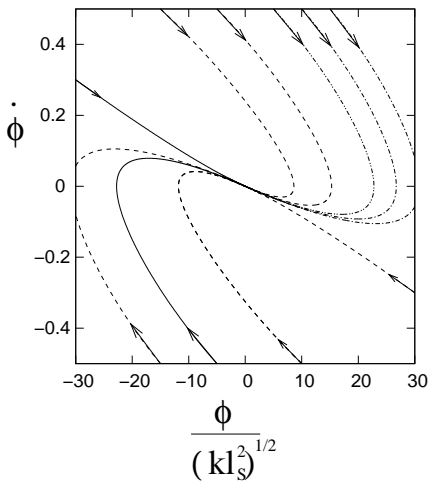}
\end{center}
\caption{Plot of the phase space solution with a variety of initial conditions.
Here we see the presence of global attractor at $(\phi=0, \dot{\phi}=0)$}
\label{phase}
\end {figure}
%%%%%%%%%%%%%%%%%%%%%%%%%%%%%%%%%%%%%%%%%%%%%%%%%%%%%%%%%%%%%%%%%%%%%%%%%%%%%%%%%%%%%%%%%%%%%%%%%%%
\begin{eqnarray}
H^2 &=& \frac{e^{Y_3}}{\sqrt{1-Y_2^2}}+\frac{\rho_B}{3M_p^2}.
\end{eqnarray}
Simple analysis shows us that critical point is at $ Y_1 =0$ and $Y_2 = 0$ which is a global attractor.
This agrees with our physical intuition since it implies the probe brane will slow down, eventually coming to
rest at the origin of the transverse space. In terms of our critical ratios we find
\begin{eqnarray}
\Omega_{\phi} &=& \frac{e^{Y_3}}{e^{Y_3}+\frac{ \rho_B}{3M^2_p}\sqrt{1-Y^2_2}} \\
\Omega_B &=& \frac{\rho_B}{\frac{3M_p^2e^{Y_3}}{\sqrt{1-Y_2^2}}+ \rho_B}
%q &=& \frac{3}{2}\gamma -1 -\frac{3\left(\gamma-Y^2_2\right)e^{Y_3}}{2\left(e^{Y_3}
%+\frac{\rho_B}{3M^2_p}\sqrt{1-Y_2^2}\right)}\\
%\lambda &=& -\frac{M_pV_{\phi}}{V^{3/2}} \\
%        &=& -\frac{1}{\sqrt{3}}e^{-Y_3/2}\frac{dY_3}{dY_1}
\end{eqnarray}

%%%%%%%%%%%%%%%%%%%%%%%%%%%%%%%%%%%%%%%%%%%%%%%%%%%%%%%%%%%%%%%%%%%%%%%%%%%%%%%%%%%%%%%%%%%%%%%%%%
%%%  Omega Vs N                                                                                 %%
%%%%%%%%%%%%%%%%%%%%%%%%%%%%%%%%%%%%%%%%%%%%%%%%%%%%%%%%%%%%%%%%%%%%%%%%%%%%%%%%%%%%%%%%%%%%%%%%%%
\begin {figure}[h]
\begin{center}
\vspace{0.5in}
\includegraphics[width=2.0in,angle=-90]{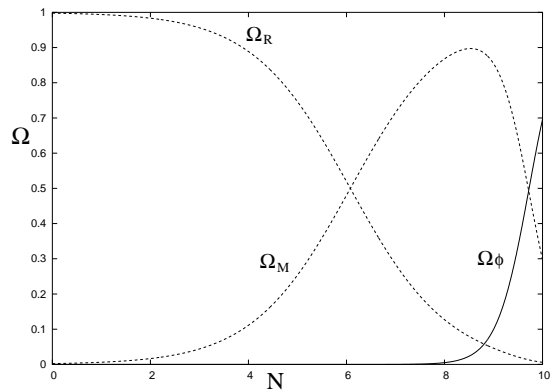}
\end{center}
\caption{Illustration of the various behaviour for $\Omega_i$. Here we have taken $\rho_m^0 = 4.58 \times10^6$, $\rho_R^0=10^{10}$ and $V_0=10^{-6}$.
The dark line is for $\Omega_R$, dotted line is for $\Omega_{\phi}$ and light line is for $\Omega_m$}
\label{omega}
\end {figure}
%%%%%%%%%%%%%%%%%%%%%%%%%%%%%%%%%%%%%%%%%%%%%%%%%%%%%%%%%%%%%%%%%%%%%%%%%%%%%%%%%%%%%%%%%%%%%%%%%%%

\noindent
Note that they are constrained by $\Omega_{\phi}+\Omega_B = 1$. We also have $\Omega_B=\Omega_M+\Omega_R$,
where $M$ and $R$ denote matter and radiation respectively, whilst $\phi$ is associated with our scalar field.\\
From the plots fig(\ref{omega}) we see that the $\Omega_{\phi}$ goes to $0.7$ and $\Omega_M$
goes to $0.3$ and $\Omega_R$ goes to $0$ in the presence epoch.
%%%%%%%%%%%%%%%%%%%%%%%%%%%%%%%%%%%%%%%%%%%%%%%%%%%%%%%%%%%%%%%%%%%%%%%%%%%%%%%%%%%%%%%%%%%%%%%%
%%% Equation of state                                                                         %%
%%%%%%%%%%%%%%%%%%%%%%%%%%%%%%%%%%%%%%%%%%%%%%%%%%%%%%%%%%%%%%%%%%%%%%%%%%%%%%%%%%%%%%%%%%%%%%%%
\begin{figure}[b]
\begin{center}
\includegraphics[width=2.0in,angle=-90]{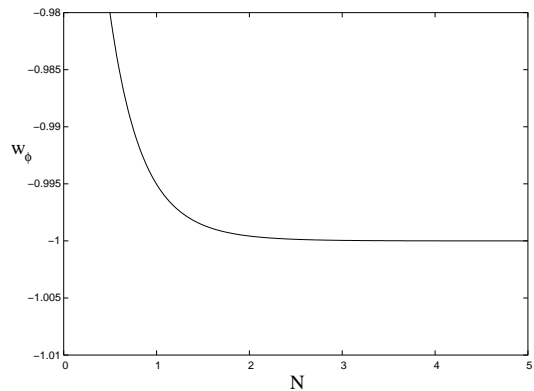}
\end{center}
\caption{Evolution of the equation of state parameter with the number of e-folds. Note that $\omega$ rapidly
approaches -1 as expected.}
\label{eos}
\end{figure}
%%%%%%%%%%%%%%%%%%%%%%%%%%%%%%%%%%%%%%%%%%%%%%%%%%%%%%%%%%%%%%%%%%%%%%%%%%%%%%%%%%%%%%%%%%%%%%%%
\noindent
We see that at late times, the field settles at the potential minimum leading to
de-Sitter solution with energy scale $V_0={\tau_3 R}/{\sqrt{kl_s^2}}$.
Using the numerical data from the preceding sections we can write this an upper bound on the energy density as follows
\begin{equation}
V_0 \le 10^{-12} M_p^4.
\end{equation}
Although this is several orders of magnitude higher than the observed value, we note that this value is
heavily dependent on the scales in the theory, and with appropriate tuning could be substantially smaller.
Since there exists no realistic scaling solution (which could
mimic matter/radiation), the model also requires the fine tuning of the initial value of the scalar field. The
field should remain sub dominant for most of the cosmic evolution and become comparable to the background at
late times. It would then evolve to dominate the background energy density ultimately settling down in the de-Sitter
phase.

However, recall from the bulk picture that the point $\sigma=0, \rho=0$ will be gravitationally unstable and
the probe brane will eventually be attracted toward the ring. In terms of our cosmological theory we see that this
de-Sitter point will actually be only quasi-stable and that a tachyonic field will eventually condense forcing the
vacuum energy down toward zero. This suggests that the vacuum energy will not be constant, but will slowly varying.
Furthermore our equation of state should be modified to incorporate the dynamics of this additional field. It is
trivial to see that the inflationary phase will terminate and give way to a dark energy phase where $\omega \sim -1$.
Once the tachyon field starts to roll, $\omega$ will increase toward zero from below giving rise to a phase of quintessence \cite{quintessence}.
Eventually we will begin to probe the strong coupling regime and our effective action will break down.

let us return to the bulk picture to understand this in more detail. We introduce a complex field $\xi = \rho + i\sigma$
which can actually be globally defined in the target space. The harmonic function factorises in this coordinate
system into holomorphic and anti-holomorphic parts $F(\xi, \bar{\xi}) = f(\xi)f(\bar{\xi})$. Thus the tachyon map will also split accordingly
\begin{equation}
\partial_t \phi = f(\xi) \partial_t \xi, \hspace{0.5cm} \partial_t \bar{\phi} = f(\bar{\xi}) \partial_t \bar{\xi}.
\end{equation}
These expressions are exactly solvable provided we continue them into the complex plane.
If we now re-construct the potential for these fields in terms of our holographic theory we obtain the general
solution
\begin{equation}\label{eq:generalpot}
V(\phi, \bar{\phi}) = \frac{R\tau_3}{\sqrt{kl_s^2}} \left\lbrack
\cos\left(\frac{\phi}{\sqrt{kl_s^2}}\right)
\cos\left(\frac{\bar{\phi}}{kl_s^2} \right) \right\rbrack^{1/2}.
\end{equation}
Clearly when $\phi$ is real we recover our $cosine$ potential, whilst if it is purely imaginary we recover the
$cosh$ solution. These correspond to motion inside the ring and motion transverse to the ring respectively.
The tachyonic instability forces the field from the false vacuum state toward the true ground state.
Therefore we expect the dark energy potential to be
\begin{equation}
V(\phi, \bar{\phi}) \sim \frac{R\tau_3}{\sqrt{kl_s^2}} \cos \left(\frac{\phi}{\sqrt{kl_s^2}} \right),
\end{equation}
and so the true minimum will occur when $V \sim 0$ at $\phi = \pm \pi \sqrt{kl_s^2}/2$ corresponding to the location
of the ring in the bulk picture. The cosmological dynamics in this particular phase are well described
by \cite{TWinf, ringinf}, where it was shown to be possible for the true vacuum to be non-zero,
provided the trajectory of the probe brane is sufficiently fine tuned.

We finally comment on the instability for the field fluctuations
for potential with a minimum \cite{FKS}. In a flat FRW background each Fourier mode of $\phi$ satisfies the
following equation
\begin{eqnarray}
&& \frac{\delta \ddot{\phi}_{\tilde{k}}}{1-\dot{\phi}^2}+
\left[3H + \frac{2\dot{\phi}\ddot{\phi}}{(1-\dot{\phi}^2)^2}\right]\delta\dot{\phi}_{\tilde{k}} \nonumber \\
&&+\Big[\frac{\tilde{k}^2}{a^2} +(\ln V)_{\phi,\phi}\Big] \delta \phi_{\tilde{k}}= 0
\end{eqnarray}
Where $\tilde{k}$ is the comoving wavenumber. We now compute the second derivatives of the potential and obtain
\begin{eqnarray}
\left(\ln V\right)_{\phi,\phi} &=&\frac{1}{kl_s^2}\left(1-\tanh\left\lbrack\frac{\phi}{\sqrt{kl_s^2}}\right\rbrack\right).
\end{eqnarray}
Here we see that $(\ln V)_{\phi,\phi}$ is never divergent
for any value of $\phi$, and is always non-negative i.e that $(\ln V)_{\phi,\phi} \in[0,1]$.
Thus we do not have any instability associated with the perturbation $\delta \phi _k$ with our potential (\ref{poten}).
This is to be contrasted with the result obtained for the open string tachyon. which has rapid fluctuations and
instabilities associated with its evolution.

\section{Conclusion}
In this note we have examined the time dependant configuration of a single $D3$ brane in the background of $NS5$ branes
distributed on a ring of radius $R$, taking the near horizon approximation.
We then studied the cosmological implications of the effective potential
which arises due to the transverse motion of $D3$ with respect to the plane of the ring.
The model appears to describe an inflationary phase giving way to a natural reheating mechanism, and then a further
phase of dark energy driven expansion. Although we cannot accurately predict the scale of the energy density at this
point, we do obtain an upper bound. In this case the dark energy phase
is a late time attractor of our model, and we predict that the vacuum energy will eventually decay to zero - although on
extremely large time-scales \footnote{However we must be careful since the DBI action will not be valid once it coalesces with the $NS$5-branes
so we must assume that it passes between the branes. This requires fine tuning of the initial trajectory which
is not realistic. This problem may be resolved by switching to the description of the model in terms of
Little String Theory \cite{lst}.}.
In fact our results will be dramatically improved by keeping the full structure of the harmonic function, because
at large distances the potential is even flatter yielding even more e-foldings of inflation.
Due to the absence of scaling solutions in our field theory, we need
to tune the initial value of the scalar field such that it can become relevant only at late times.
With these described fine tunings, the geometrical field is
a potential candidate for dark energy.
The model is free from tachyon instabilities, and the field perturbations behave in a similar manner to those of
the canonical scalar field.

Of the model we have several potential problems. Firstly our assumption about the coupling of the DBI to four-dimensional
gravity, although as we have pointed out this can be resolved by a full string theory compactification. However
there will generally be large corrections, potentially destroying the simplicity of the solution.
Secondly the trajectory of the brane in the bulk space is particularly special. In the most generic case we would anticipate
a general spiralling trajectory toward the ring. In this case there would be no simple decoupling of the modes and
we would need to consider the full form of the potential. This amounts to a certain amount of fine tuning of the
initial conditions. Another problem is that we have not turned on any standard model fields which would be expected
to couple to the inflaton on the world-volume. However the inclusion of U(1) gauge fields on the brane will act to
reduce the velocity of the field by a factor of $\sqrt{1-E^2}$, where $E$ is our dimensionless electric field.
More importantly however is that we have neglected the induced two-form field strength, which can have important
applications in cosmology as seen in the recent paper \cite{bfield}.
Despite these problems, we know there is a coset model describing the background which opens the way for exact
string theory calculations. Furthermore the relationship between the two energy scales in the theory means it is
possible to talk about long-standing problems such as the Transplanckian issue \cite{trans}.
One further problem is the termination of reheating in this model. We have emphasised that this is indeed difficult
to tackle in this model due to its analytic simplicity. One may hope that a careful analysis of the tachyon mapping
will lead to more realistic behaviour for the inflaton field, and thus a possible exit from reheating. In fact this may also be possible
by considering more general trajectories of the probe brane in the bulk picture. We hope to return to this issue in a future publication.

One thing that emerges though is the relationship between a dark energy dominated phase and the 'fast rolling' DBI action \cite{dbiinflation}.
Although our proposal is far from rigorous, it does capture the majority of the same physics as in the
flux compactification scenario. We know that $D$-branes moving in non-trivial backgrounds have sub-luminal velocities
as measured by observers in the far UV of the geometry, due to the gravitational red-shifting. In fact the
branes are decelerating and for late times will have negligible velocities. This in turn implies that the
equation of state parameter will tend to $\omega \sim -1$ at late times. A concrete example where this could be examined
is in the case of the warped deformed conifold \cite{conifold}. The RR flux will wrap the $S^3$ in the IR end of the geometry, and we can
imagine a solitary $D3$-brane probing this part of the conifold after an inflationary phase. To an observer in
the compact space the brane will slow down as it reaches the origin of the $S^3$ yielding a dark matter dominated
phase \cite{darkenergy}.

However our model opens up the possibility that non-trivial background configurations may have important
implications for brane cosmology, as we have seen how to combine inflation, reheating and dark energy in a single
model. Furthermore this is not subject to the same landscape problems as the flux compactification models, and
we can try and tackle higher energy issues in a clear formalism \cite{warpcosm}.
Although we acknowledge the simplicity of our solution we hope that this will encourage more research
in this direction.

\section{Acknowledgement}
We thank M. Sami, S. Tsujikawa, S. Thomas for discussion and
critical comments. BG thank D. Samart and S. Pantian for
re-checking numerical plots and calculation. BG is supported by
the Thailand Research Fund. Tapan Naskar thanks Jamia for
hospitality. JW is supported by a Queen Mary studentship.
%%%%%%%%%%%%%%%%%%%%%%%%%%%%%%%%%%%%%%%%%%%%%%%%%%%%%%%%%%%%%%%%%%%%%%%%%%%%%%%%%%%%%%%%%%%%%%%%%%%

\end{document}